\documentclass[fleqn,10pt,a4paper]{SelfArx} 

\usepackage[english]{babel} 

\usepackage{lipsum} 
\usepackage{multirow}
\usepackage{booktabs}
\usepackage{xcolor}
\usepackage{balance}

\definecolor{color1}{RGB}{0,0,90} 
\definecolor{color2}{RGB}{0,20,20} 

\usepackage{hyperref} 
\hypersetup{hidelinks,colorlinks,breaklinks=true,urlcolor=color2,citecolor=color1,linkcolor=color1,bookmarksopen=false,pdftitle={Title},pdfauthor={Author}}
\usepackage{xspace}

\JournalInfo{\includegraphics[width=3cm]{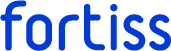}} 
\Archive{fortiss GmbH\\ White Paper, 2018} 

\PaperTitle{Model-Based Safety and Security Engineering} 

\Authors{Vivek Nigam\textsuperscript{1}*, Alexander Pretschner\textsuperscript{1,2}, and Harald Ruess\textsuperscript{1}} 
\affiliation{\textsuperscript{1}\textit{fortiss GmbH, Munich, Germany}} 
\affiliation{\textsuperscript{2}\textit{Technical Univeristy of Munich, Munich, Germany}} 
\affiliation{*\textbf{Corresponding author}: nigam@fortiss.org} 

\newcommand{\keywordname}{Keywords} 

\newcommand\ie{\emph{i.e.}\xspace}
\newcommand\eg{\emph{e.g.}\xspace}
\newcommand\etal{\emph{et al.}\xspace}

\Abstract{
By exploiting the increasing surface attack of systems,  cyber-attacks can cause catastrophic events, such as, remotely disable safety mechanisms. This means that in order to avoid hazards, safety and security  need to be integrated, exchanging information, such as, key hazards/threats, risk evaluations, mechanisms used. This white paper describes some steps towards this integration by using models. We start by identifying some key technical challenges. Then we demonstrate how models, such as Goal Structured Notation (GSN) for safety and Attack Defense Trees (ADT) for security, can address these challenges. In particular,  (1) we demonstrate how to extract in an automated fashion security relevant information from safety assessments by translating GSN-Models into ADTs; (2) We show how security results can impact the confidence of safety assessments; (3) We propose a collaborative development process where safety and security assessments are built by incrementally taking into account safety and security analysis; (4) We describe how to carry out trade-off analysis in an automated fashion, such as identifying when safety and security arguments contradict each other and how to solve such contradictions. We conclude pointing out that these are the first steps towards a wide range of techniques to support Safety and Security Engineering. As a white paper, we avoid being too technical, preferring to illustrate features by using examples and thus being more accessible.
}
\begin{document}

\flushbottom 

\maketitle 

\tableofcontents 

\thispagestyle{empty} 


\section*{Introduction} 
\addcontentsline{toc}{section}{Introduction} 

The past years have witnessed a technological revolution interconnecting systems and people. This revolution is leading to new exciting services and business models. Managers can now remotely adapt manufacturing to customers needs in the so called Industry 4.0 paradigm. In the very near future, vehicles will operate with high levels of autonomy making decisions based on information exchanged with other vehicles and the available infrastructure. Similarly, autonomous UAVs will be used to transport cargo and people.

This technological revolution, however, leads to new challenges for safety and security. \textbf{The greater connectivity of systems increases their attack surface}~\cite{attack.surface}, that is, the ways an intruder can carry out an attack. Whereas in conventional systems, attacks, such as car theft, require some proximity to their target, with interconnected systems, cyber-attacks, such as stealing sensitive data or hacking into safety-critical components, can be carried out remotely.

This increased attack surface has important consequences for system safety. Security can no longer be taken lightly when assessing the safety of a system.\footnote{By, for example, enforcing that only authorized persons are near to sensitive parts of the system.} \textbf{Indeed, cyber-attacks can lead to catastrophic events}~\cite{attack.mill,attack.cars,attack.jeep,attack.airplane}. For example, cyber-attacks exploiting a connection to an autonomous vehicle may remotely disable safety features, such as airbags, thus placing passengers in danger~\cite{attack.cars}. \textbf{Therefore, both safety and security have to be taken into account in order to conclude the safety of a system.}

On the other hand, this increased surface attack also has important consequences to security itself. While traditional security concerns, such as, handling physical attacks, \eg, car theft, remain important concerns, security engineers shall consider a wider range of 
cyber-attacks due to the increased attack surface and, in particular, cyber-attacks that lead to catastrophic events. This means that security engineers shall understand information normally contained in safety assessments, such as, which are the catastrophic events and how they can be caused/triggered. That is, \textbf{security analysis shall take into account safety concerns.} 

Finally, \textbf{this technological revolution will also have an impact on system certification processes.} It is unreasonable to allow the delivery of products to consumers without considering attacks that may lead to catastrophic events. Companies will soon need to provide detailed security analysis arguing that the risk of such threats is acceptable. While current certification agencies do not demand such analysis, there have been initiatives towards this direction, \eg, RTCA DO-326A~\cite{do326}, SAE J3061~\cite{saeJ3061}, ISO 21434~\cite{iso21434}, ISO 15408~\cite{iso15408}. \textbf{This change on certification processes will have an important impact to the processes and business models of companies.} It is, therefore, important to develop techniques that integrate safety and security that can facilitate the development of such safety and security arguments.

\paragraph{Problem Statement} Safety and security are carried out with very different mind-sets. While safety is concerned in controlling catastrophic events caused by the malfunctioning of the system, security is concerned in mitigating attacks from malicious agents to the system. 

The difference between safety and security is reflected in the types of techniques used for establishing safety and security assessments. Safety assessments are constructed by using analysis techniques such as Fault Tree Analysis (FTA), Failure Mode and Effect Analysis (FMEA), System Theoretic Process Analysis (STPA), Goal Structured Notation~\cite{gsn11standard}, specific safety designs and mechanisms, \eg, Voters, Watchdogs, etc. Security, on the other hand, uses different assessment techniques, such as Attack Trees~\cite{schneier99jst}, Attack Defense Trees~\cite{kordy10fast}, STRIDE, and security designs and mechanisms, \eg, access control policies, integrity checks, etc. 

It is not clear how these different techniques can be used in conjunction to build a general safety and security argument. Questions such as the following have to be answered:
\textbf{What is the common language for safety and security?
How can security engineers use safety assessments to build security arguments? What is the impact of security assessments to safety cases?
What are the trade-offs to be considered? Which methods can be implemented within current practices?} 

This difference in mentality is also reflected in the development process leading many times to \textbf{Poor Process Integration of Safety and Security}. Safety and security concerns are only integrated at very late stages of development when fewer solutions for conflicts are possible/available. For example, in secure by design processes, security engineers participate from the beginning proposing security design requirements. However, they do not take into account the impacts of such requirements on safety arguments, for example, adding mechanisms with unreasonable delays. \textbf{The lack of Integration of Safety and Security leads to increased development costs, delays and less safe and secure systems.}

\paragraph{Benefits of Safety and Security Integration} Besides improving the safety and the security of systems, the integration of safety and security can lead to a number of benefits. We highlight some possible benefits:

\begin{itemize}
  \item \textbf{Early-On Integration of Safety and Security:} Safety and security assessments can be carried out while the requirements of system features are established. Safety assessments provide concrete hazards which should be treated by security assessments, thus \textbf{helping security engineers to set priorities}. For example, a safety hazard shall be given a higher priority compared to other security attacks which do not cause catastrophic events;

  \item \textbf{Verification and Validation:} While safety has many well-established methods for verification, security verification relies mostly on penetration testing techniques, which are system dependent and therefore, resource intensive. The integration of Safety and Security can facilitate security verification. \textbf{Much of knowledge gathering can be retrieved from safety assessment, thus saving resources}. For example, FTAs describe the events leading to some hazardous event, while FMEAs describe single-points of failures. This information can be used by security engineers to plan penetration tests, \eg, exploit single-point of failures described in FMEAs, thus leading to increased synergies and less development efforts;

    \item \textbf{Safety and Security Mechanisms Trade-Off Analysis:} By integrating safety and security analysis, it is possible to analyze trade-offs between control and counter-measures proposed to support safety and security arguments. On the one hand, \textbf{safety and security measures may support each other, making one of them superfluous.} For example~\cite{glas14safesec,novak07etfa}, there is not need to use CRC (Cyclic Redundancy Check) mechanisms for safety, if messages are being signed with MAC (Message Authentication Codes) as the latter already achieves the goal of checking for message corruption. On the other hand, \textbf{safety and security mechanisms may conflict with each other.} For example, emergency doors increase safety by allowing one to exit a building in case of fire, but it may decrease security by allowing unauthorized persons to enter the building. Such trade-off analysis can help solve conflicts as well as identify and remove redundancies reducing product and development costs.
\end{itemize}

\paragraph{Outline} Model-Based Engineering (MBE) is widely used by industries such as automotive~\cite{iso26262}, avionics~\cite{do331} and Industry 4.0~\cite{IEC61499} for developing systems. The scope of the methods we propose will assume a MBE development where models play a central role. \textbf{As a white paper, we will avoid being too deep (and technical), preferring to illustrate the range of possibilities with examples.} In future works, however, we will describe in detail the techniques used as well as propose extensions.

Section~\ref{sec:challenges} identifies key technical challenges for the integration of safety and security. Section~\ref{sec:techniques} reviews briefly the main techniques used for safety and security. Section~\ref{sec:MBE} describes how MBE provides the basic machinery for integrating safety and security. Section~\ref{sec:saf2sec} describes, by example, how one can extract security relevant information from safety assessments. Section~\ref{sec:sec2saf} describes how the evaluation of security assessments can impact the confidence in safety assessments. Section~\ref{sec:collaboration} builds on the material in the previous sections and proposes a \emph{collaborative safety and security development process}. 
Section~\ref{sec:trade-off} describes how to carry out trade-off analysis between safety and security mechanisms. We illustrate how the detection of conflicts can be done automatically. Finally, Section~\ref{sec:steps} concludes by pointing out our next steps.

\textbf{We also point out that the techniques described here have been implemented or under implementation as new features of the Model-Based tool AF3~\cite{af3} maintained by fortiss's MBSE department.}


\section{Main Technical Challenges}
\label{sec:challenges}
This section introduces some of challenges that we believe are important for safety and security integration and therefore, shall and will be tackled in the following years.

In the Introduction, we mentioned that the difference in safety and security mind-sets lead to different techniques for carrying out safety and security assessments. One lacks a common ground, that is, a language that can be used to integrate both safety and security assessments. Without such common ground there is little hope to integrate safety and security in any meaningful way. This leads to our first challenge:

\begin{center}
  \textbf{Challenge 1:} \emph{Develop a common language which can be used to integrated safety and security assessments.}
\end{center}

Safety assessments contain useful information for security engineers to evaluate how attacks can cause catastrophic events. Indeed, safety assessments contain the main hazards, how these can be triggered, control mechanisms installed, entry points, etc. However, safety assessments are written in the most varied forms and often in non-machine readable formats, \eg, Word documents. This prevents security engineers to use safety assessments effectively in order to understand how cyber-attacks could affect the safety of a system. This leads to our second challenge:

\begin{center}
  \textbf{Challenge 2:} \emph{Develop techniques allowing the (semi-)automated extraction of relevant security information from safety assessments.}
\end{center}

It is not reasonable to conclude the safety of an item without considering its security concerns. For example, an airbag cannot be considered safe if an intruder can \emph{easily} deploy it at any time. This means that security assessments related to safety hazards shall impact security assessments. So, if a security assessment concludes that there is unacceptable security risk of the airbag being deployed, this conclusion should render the airbag safety unacceptable. To do so, we need techniques for incorporating the conclusions of the security assessments in safety assessments.
This leads to our third challenge:

\begin{center}
  \textbf{Challenge 3:} \emph{Develop techniques allowing the incorporation of relevant security assessment findings into safety assessments.}
\end{center}

Safety and security assessments often lead to changes on the architecture by, for example, incorporating control and mitigation mechanisms. Many times, these mechanisms may support each other to the point of being redundant. If this is the case, some mechanisms may be removed thus reducing costs. On the other hand, some mechanisms may conflict each other. Therefore, decisions may need to be taken, \eg, finding alternative mechanisms or prioritizing safety over security or vice-versa. That is, trade-offs shall be carried out. However, there are no techniques for carrying out such trade-offs, leading to our fourth challenge:

\begin{center}
  \textbf{Challenge 4:} \emph{Develop techniques allowing the identification of when safety and security mechanisms support, conflict or have no effect on each other, and to carry out trade-off analysis.}
\end{center}

In order to certify a system, developers have to provide enough evidence, \ie, safety arguments, supporting system safety. As mentioned in the Introduction, with interconnected systems, safety arguments shall also consider security. Currently, safety arguments provide detailed quantitative evaluation for the safety of systems based on the probability faults. When taking security into account such quantitative evaluations no longer make sense as the probabilities of an attack to occur are in another order of magnitude as the probability of faults. Unfortunately, there are no techniques to present such quantitative safety arguments taking into account security. This leads to our last challenge:

\begin{center}
  \textbf{Challenge 5:} \emph{Develop techniques for quantitative evaluation of system safety including acceptable risk from security threats.}
\end{center}

This white paper provides some ideas on how we plan to tackle these challenge. Challenge 1 and 2 are treated in Section~\ref{sec:saf2sec}; Challenge 3 and 5 are treated in Section~\ref{sec:sec2saf}. Challenge 4 is treated in Section~\ref{sec:trade-off}. We plan in the next years to build and expand on these ideas.

\section{Safety and Security Techniques}
\label{sec:techniques}
This section briefly reviews the main techniques used for establishment of safety and security as well as some proposals for integrating safety and security. 
Our goal is not to be comprehensive, but rather review established techniques that will be used in the subsequent sections. 

\subsection{Safety}
\label{subsec:safety}
We review some techniques used by engineers to evaluate and increase the safety of a system, namely, Fault Tree Analysis (FTA), Failure Modes and Effect Analysis (FMEA),  Goal Structured Notation, and safety mechanisms.

\paragraph{Fault Tree Analysis (FTA):} FTA is a top-down approach used in order to understand which events may lead to undesired events. It is one of the most important techniques used in safety engineering. An FTA is a tree with the root labeled with the top undesired event. The tree contains ``and'' and ``or'' nodes specifying the combination of events that can lead to the undesired event.

\begin{figure}
\begin{center}
\includegraphics[width=0.8\columnwidth]{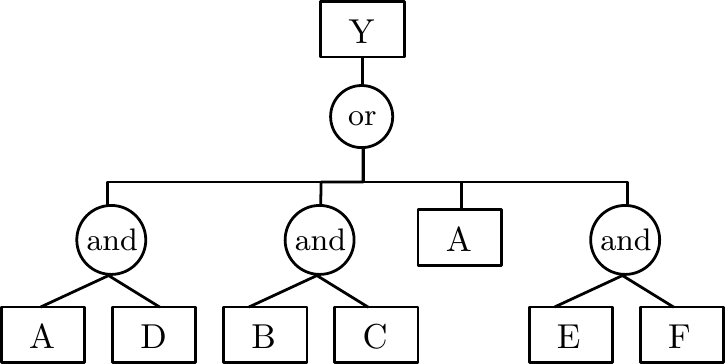}  
\end{center}
\caption{FTA Example.}
\label{fig:fta}
\end{figure}

Consider, for example, the FTA depicted in Figure~\ref{fig:fta}. The undesired event is Y placed at the root of the tree. A safety engineer is interested on the \emph{cut sets of an FTA}, that is, the collections of events, normally component faults, that lead to the undesired event. For this FTA example, the cut sets are:
\begin{center}
  \{A,D\}, \{B,C\}, \{A\}, \{E,F\}
\end{center}
as any of these combinations lead to the event Y. If both A and D happen at the same time, the left-most and branch is satisfied leading to the event Y. 

From a FTA, one can compute the \emph{minimum cut sets}, that is, the minimum set of cut sets that encompasses all possible combinations of triggering an undesired event. The minimum cut set for the given example is 
\begin{center}
  \{B,C\}, \{A\}, \{E,F\}
\end{center}
Notice that the event A already triggers the event Y. Therefore, there is no need to consider the cut set \{A,D\}as it is subsumed by the cut set \{A\}.

Given the minimum cut sets, a safety engineer can, for example, show compliance with respect to the safety requirements. This may require placing control measures to reduce the probability of the corresponding undesired event.

As we argue in Section~\ref{sec:saf2sec}, FTA provides very useful information for security engineers. Indeed, an attack triggering, for example, the event A would lead to an undesired (possibly catastrophic) event. This means that penetration tests could be more focused in assessing how likely/easy it is to trigger event A, rather than finding this from scratch.

\paragraph{Failure Modes and Effect Analysis (FMEA):} FMEA is a bottom-up approach (inductive method) used to systematically identify potential failures and their potential effects and causes. Thus FMEA complements FTA by instead of reasoning from top-level undesired events as in FTA, adopting a bottom-up approach by starting from faults/failures of sub-components to establish top level failures. 


FMEAs are, normally, organized in a table containing the columns: Function, Failure Mode, Effect, Cause, Severity, Occurrence, Detection and the RPN value.

Failure modes are established for each function. Examples of failure modes include~\cite{ar4761}:
\begin{itemize}
  \item \textbf{Loss of Function}, that is, when the function is completely lost;
  \item \textbf{Erroneous}, that is, when the function does not behave as expected due to, for example, an implementation bug;
  \item \textbf{Unintended Action}, that is, the function takes the action which was not intended;
  \item \textbf{Partial Loss of Function}, that is, when the function does not operate at full operation, \eg, some of the redundant components of the function are not operational.
\end{itemize}

Effect and cause are descriptions of, respectively, the impact of the failure mode of the function to safety and what could a cause for such failure be, \eg, failures of sub-components. Severity, Occurrence and Detection are numbers, ranging normally from 1-10. The higher the value for severity the higher the impact of the failure. The higher the value for occurrence the higher is the likelihood of the failure. The higher the value of detection the less likely it is to observe (and consequently activate control mechanisms) the failure. 

Finally, the value RPN is computed by multiplying the values for severity, occurrence and detection. It provides a quantitative way of classifying the importance of failure modes. The higher the value of RPN of a failure the higher its importance.

As we argue in Section~\ref{sec:saf2sec}, FMEAs also provide useful information for security engineers. For example, it describes the severity of failure modes and its causes. Therefore, security engineers can use this information to prioritize which attacks to consider. Notice, however, that occurrence does not play much importance for security engineers as occurrence in FMEA does not reflect the likelihood of attacks to occur, but rather the likelihood of faults/failures.


\paragraph{Safety Mechanisms:} Safety mechanisms, such as voters, watchdogs, are often deployed in order to increase the safety of a system. For example, consider the hazard \emph{unintended airbag deployment}. Instead of relying on a single signal, \eg, crashing sensor, to deploy an airbag, a voter can be used to decide to deploy an airbag taking into account multiple (independent) signals, \eg, crashing sensor and gyroscope, thus reducing the chances for this hazard. 

However, as pointed out by Preschern \etal~\cite{preschern13plop}, safety mechanisms themselves can be subject to attacks. For example, an attacker may tamper the voter leading to a hazard. As we detail in Section~\ref{sec:saf2sec}, if security engineers are aware of the deployment of such mechanisms, they can assess how likely it is to attack them to trigger a hazard.

\paragraph{Goal Structured Notation (GSN):} Safety assessments are complex, breaking an item safety goal into safety sub-goals, \eg, considering different hazards, and often applying different methods, \eg, FTA, FMEA, Safety Mechanisms. GSN~\cite{gsn11standard} is a formalism introduced to present safety assessments in a semi-formal fashion. 

\begin{figure}[t]
\begin{center}
  \includegraphics[width=0.99\columnwidth]{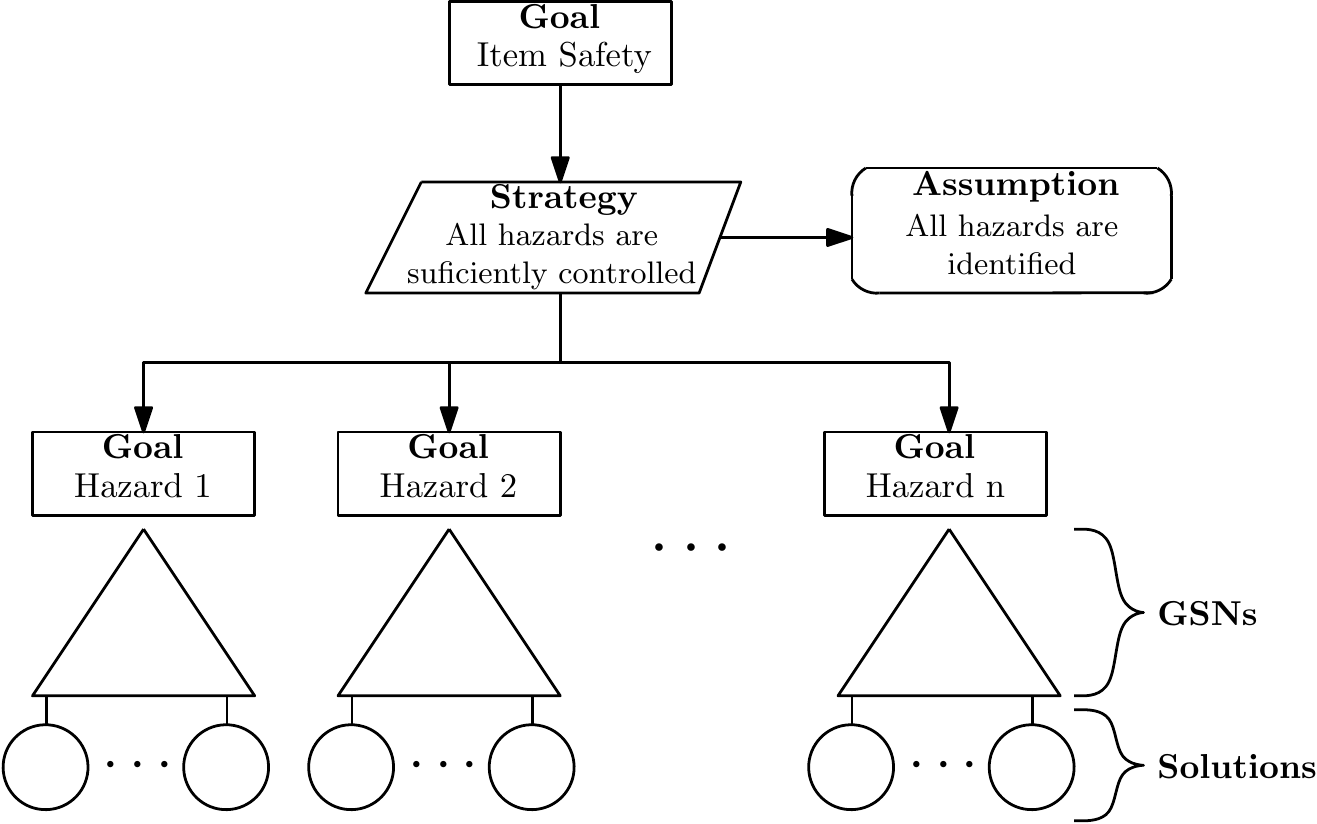}  
\end{center}
\caption{GSN Hazard Pattern.}
\label{fig:gsn-haz}
\end{figure}

\begin{figure}[t]
  \begin{center}
  \includegraphics[width=0.85\columnwidth]{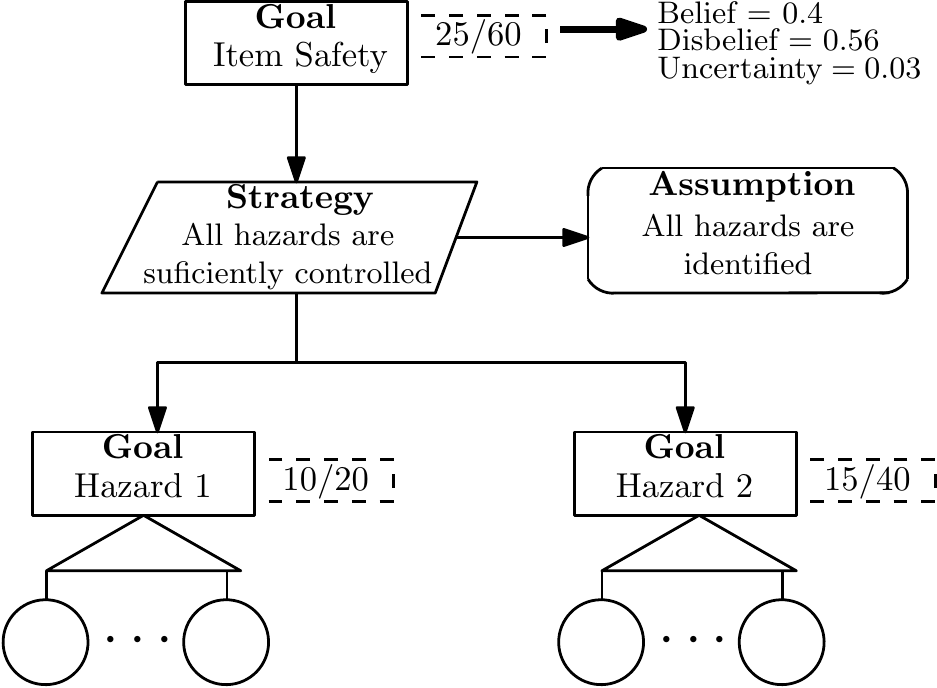}  
\end{center}
\caption{Example of GSN-Model with Quantitative Information. Here the pair $m/n$ attached to goals specifies, respectively, the number of defeaters outruled and the total number of identified defeaters.}
\label{fig:gsn-quantitative}
\end{figure}

Since its introduction, different safety arguments have been mapped to GSN patterns. Consider, for example, the GSN pattern depicted in Figure~\ref{fig:gsn-haz}. It specifies the argument by analysing all the possible/known hazards to an item's safety. It is assumed that all the hazards are known. For each hazard a safety argument, also represented by a GSN-Model, is specified. At the leaves of the GSN-Model, one describes the \emph{solutions} that have been taken, \eg, carry out FTA, FMEA, safety designs, etc. 

Clearly, such safety arguments can provide important information for security. For example, it contains the key safety hazards of an item. It also contains what type of solutions and analysis have been carried out. However, a problem of GSN-Models is the lack of more precise semantics. The semantics is basically the text contained in the GSN-Models, which may be enough for a human to understand, but it does not provide enough structure for extracting automatically security-relevant information. In Section~\ref{sec:saf2sec}, we extend GSN-Models and show how to construct security models, namely, Attack Trees, from a GSN-Model.

Finally, recent works~\cite{wang17safecomp,duan16hases} have proposed mechanisms for associating GSN-Models with quantitative values denoting its confidence level. These values are inspired by Dempster-Shafer Theories~\cite{demptser-schaffer} containing three values for, respectively, the Belief, Disbelief, and Uncertainty on the safety assessment. These values may be assigned by safety experts~\cite{wang17safecomp} or be computed from the total number of identified defeaters\footnote{A defeater is a belief that may invalidate an argument.} and the number of defeaters one was able to outrule~\cite{duan16hases}. 

We illustrate the approach proposed by Duan \etal~\cite{duan16hases}. Consider the GSN-Model depicted in Figure~\ref{fig:gsn-quantitative}. It contains a main goal which is broken down into two sub-goals. GSN goals are annotated with the number of defeaters outruled and the total number of defeaters. Intuitively, the greater the total number of defeaters, the lower is the uncertainty. Moreover, the greater the number of outruled defeaters the greater the belief on the GSN-Model and the lower the disbelief. In Figure~\ref{fig:gsn-quantitative}, a total of 60 = 20 + 40 defeaters have been identified and only 25 = 10 + 15 have been outruled. These values yield a Belief of 0.4, Disbelief of 0.56 and Uncertainty of 0.03.\footnote{We refer to the work of J{\o}sang~\cite{josang01} on how exactly these values are computed.} If further 20 defeaters are outruled, then the Belief is increased to 0.73, the Disbelief reduces to 0.24 and the Uncertainty remains the same value 0.03.

Intuitively, only arguments that have high belief, thus low uncertainty and low disbelief, shall be accepted. As we argue in Section~\ref{sec:sec2saf}, such a quantitative information can be used to incorporate the results of security assessments in safety assessments. For example, if no security assessment has been carried out for a particular item, then the associated uncertainty shall increase. On the other hand, if a security has been carried out establishing that the item is secure, then the belief on the safety of the item shall increase. Otherwise, if an attack is found that could compromise the safety of the item, then the disbelief shall increase.

\subsection{Security}
\label{subsec:security}
We review some models used for carrying out threat analysis. More details can be found in Shostack's book~\cite{shostack14book} and in the papers cited.

\paragraph{Attack Trees:} First proposed by Schneier~\cite{schneier99jst}, attack trees and its extensions~\cite{bistarelli06ares,kordy10fast} are among the main security methods for carrying out threat analysis. An attack tree specifies how an attacker could pose a threat to a system. It is analogous to GSN-Models~\cite{gsn11standard} but, instead of arguing for the safety of a system, an attack tree breaks down the possibilities of how an attack could be carried out.  

\begin{figure}
  \begin{center}
    \includegraphics[width=0.7\columnwidth]{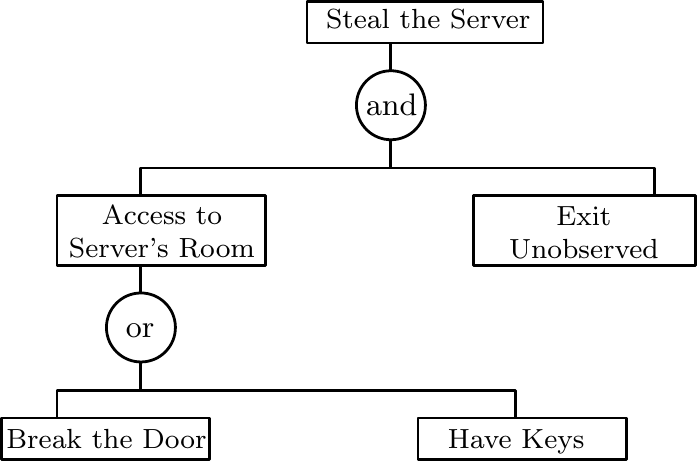}
  \end{center}
  \caption{Attack Tree Example.}
  \label{fig:attack-tree}
\end{figure}

Consider, for example, the Attack Tree depicted in Figure~\ref{fig:attack-tree}. It describes how an intruder can successfully steal a server. He needs to have access to the server's room and be able to exit the building without being noticed. Moreover, n order to access to the server's room, he can break the door or obtain the keys. 

\begin{figure}
  \begin{center}
    \includegraphics[width=0.99\columnwidth]{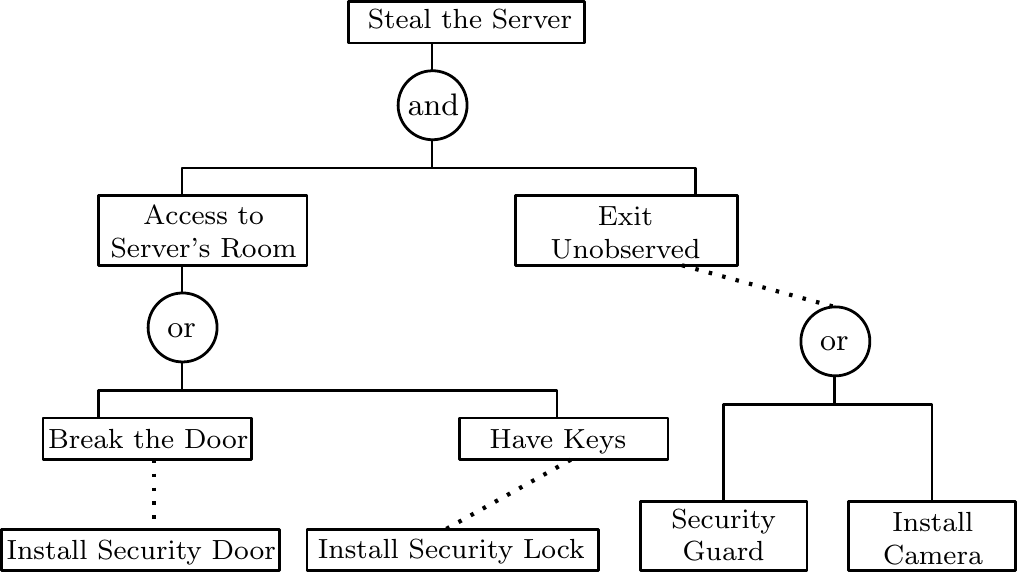}
  \end{center}
\caption{Attack Defense Tree Example.}
\label{fig:attack-defense-tree}
\end{figure}

\paragraph{Attack Defense Trees (ADTs):} Attack defense trees~\cite{kordy10fast} extend attack trees by allowing to include counter-measures to attack trees. Consider the attack defense tree depicted in Figure~\ref{fig:attack-defense-tree} which extends the attack tree depicted in Figure~\ref{fig:attack-tree}. It specifies counter-measures, represented by the dotted edges, to the possible attacks. For example, ``breaking the door'' can be mitigated by installing a security door which is harder to break into. Similarly, installing a security camera or hiring a security guard can mitigate that the attacker leaves the building undetected. Attack defense trees also allow to model how attackers could attack mitigation mechanisms. For example, a cyber-attack on the security camera causing it to record the same image reduces the camera's effectiveness. 

\paragraph{Quantitative Attack Defense Trees} Attack Defense Trees are not only a qualitative threat analysis technique, but they provide quantitative information~\cite{bagnato12ijsse,bistarelli06ares}. Quantitative information can represent, for example, ``what is the likelihood and impact of the attack?'', ``what are the costs of attacking a system?'',  ``how long does the attack require?''. Bagnato \etal~\cite{bagnato12ijsse} propose mechanisms to associate quantitative information to attack defense trees and to carry out computation to answer such questions. From the quantitative information, security engineers shall decide whether the security risk is acceptable. 



\subsection{Safety and Security}
\label{subsec:safety-security}
The problem of safety and security has been known for some time already and techniques have been proposed. They fall into the following two main categories: 

\paragraph{General Models for Both Safety and Security:} A number of works~\cite{nostro14issrew,safsec.book,meland14ijsse} have proposed using general models encompassing both safety and security concerns. For example, GSN extensions with security features, so that in a single framework, one can express both security and safety~\cite{meland14ijsse}. 

Although it is an appealing approach, it does not take into account the different mind-sets between safety and security, which poses serious doubts on the practicality of such approach. On the one hand, security engineers do use GSNs for threat modeling and it is hard to expect them to combine security threats with solutions such as FTA, FMEA, etc. On the other hand, safety engineers are not security experts, so it is hard to expect that they would develop deep security analysis. 


\paragraph{Safety Assessments used for Security Analysis:} Instead of building a general model for both safety and security, some approaches~\cite{durrwang17safecomp,sabaliauskaite18ijas,taguchi15} propose the development of safety assessments and then ``passing the ball'' to security engineers to carry out security analysis based on the safety assessments. 

An example of this approach is the use of standard (natural) language, such as Guide Words~\cite{durrwang17safecomp}, with information in safety assessments relevant for carrying out security assessments.  For example, HAZOP uses guide words to systematically describe the hazards, such as under which condition it may occur. This information can provide hints for carrying out security analysis. 

Recently, D\"urrwang \etal~\cite{durrwang17safecomp} have proposed the following set of guide words for embedded systems: \emph{disclosure}, \emph{disconnected}, \emph{delay}, \emph{deletion}, \emph{stopping}, \emph{denial}, \emph{trigger}, \emph{insertion}, \emph{reset}, and \emph{manipulation}. These words provide a suitable interface between safety and security terminology thus allowing security engineers to better understand and re-use work carried out by safety engineers. This methodology has been used~\cite{durrwang18sae} to find a security flaw in airbag devices.


\section{Integrating Safety and Security using MBE}
\label{sec:MBE}
Model-Based Engineering (MBE) proposes development by using (domain-specific) models, such as GSNs~\cite{gsn11standard}, Attack Defense Trees~\cite{schneier99jst}, Matlab Simulink~\cite{matlab}, SysML~\cite{sysml} and AF3~\cite{af3}. These models are used to exchange information between engineers to, for example, further refine requirements, implement software/systems/architecture, including software deployment. 

\begin{figure}
\begin{center}
  \includegraphics[width=0.99\columnwidth]{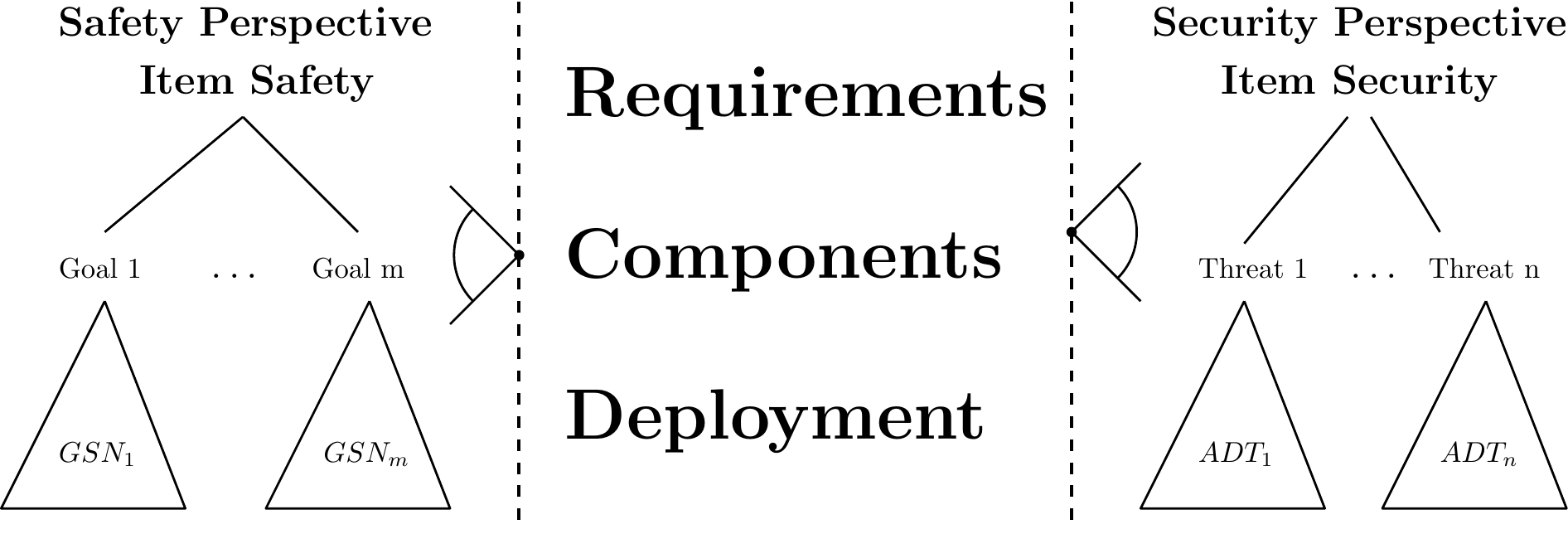}  
\end{center}
\caption{Safety and Security Lenses.}
\label{fig:safsec-lenses}
\end{figure}

This is illustrated by Figure~\ref{fig:safsec-lenses}. Requirements are traced to components that are then embedded into the given hardware. These model-elements (requirements, components and deployments) reflect different development aspects, including safety an security. Safety arguments expressed as GSN are reflected into the model-elements. For example, the handling of hazards shall be expressed as \emph{safety requirements} and safety designs, such as voters, as \emph{safety design requirements}. Similarly, threats identified by security arguments shall yield \emph{security requirements} and counter-measures as \emph{security design requirements}.


\paragraph{Features of our Approach:} As we will illustrate in the following sections, MBE provides a general framework for the integration safety and security through model-elements.
We enumerate below some of the differences/features of our approach with respect to existing approaches described in Section~\ref{subsec:safety-security}: 
\begin{enumerate}
  \item \textbf{GSN and ADT Integration:} Instead of natural language as in a Guide Words approach (see Section~\ref{subsec:safety-security}), we use models, GSN-Models and ADTs. 
Models contain much more information than Guide Words, \eg, traces to components, logical relation of solutions and hazards, quantitative evaluation. Furthermore, as we describe in Section~\ref{sec:saf2sec} and \ref{sec:sec2saf}, information from a GSN-Model can be used to construct ADTs automatically and evaluations of ADTs can be incorporated into the evaluation of GSN-Models impacting a GSN-Model's confidence;

\item \textbf{Development as a Game:} On the one hand, models that contain both safety and security annotations, like security extensions of GSN~\cite{nostro14issrew}, require that safety and security work closely together in a single model, instead of using specialized techniques and models for safety and security. On the other hand, Guide Words allow safety and security to use their own techniques, but collaboration resumes to a single ``passing the ball'' from safety to security. This means that security is not taken into account for safety. 

Our development proposal has the advantages of both methods above. It is a collaborative process where the ``ball is passed'' between safety and security engineers until an acceptable risk is reached. Moreover, it also allow safety and security engineers to use their own specialized models (GSN-Models, FTA, FMEA, etc for safety and Attack Defense Trees for security). We describe our process in Section~\ref{sec:collaboration} being illustrated by Figure~\ref{fig:collaborative}

\item \textbf{Trade-Off Analysis:} Models also help to carry out trade-off analysis. In particular, GSN-Models contain solutions, such as safety mechanisms, and ADTs contain counter-measures, such as security mechanisms (such as counter-measures). As we illustrate in Section~\ref{sec:trade-off}, we can identify when safety and security mechanisms contradict each other. Once a conflict is identified, compromises should be found, \eg, finding other mechanisms or prioritizing safety over security. We describe how such contradiction can be solved.
\end{enumerate}

\section{Safety to Security}
\label{sec:saf2sec}
This section describes how safety assessments, expressed as GSN-Models, can be used by security engineers. As described in Section~\ref{subsec:safety}, a GSN-Model contains safety details, such as the key hazards, safety methods (FTA, FMEA), and safety mechanisms used (Voters, Watchdogs) to control hazards. These details can be very useful for carrying out security assessments, such as understand which are the hazards, how they can be triggered, which safety mechanisms could be attacked. Our main goal here is to illustrate how a GSN-Model can be transformed into Attack Tree specifying a preliminary security assessment for the item assessed by the GSN-Model. 

However, the first obstacle we face is that GSN-Models are syntactic objects, where its nodes are described with (arbitrary) text, lacking thus more precise semantics. It is, therefore, not possible to extract systematically from a GSN-Model security relevant information. That is, GSN lacks a common language for safety and security integration (Challenge 1). 

\begin{figure*} 
\begin{center}
  \includegraphics[width=0.98\textwidth]{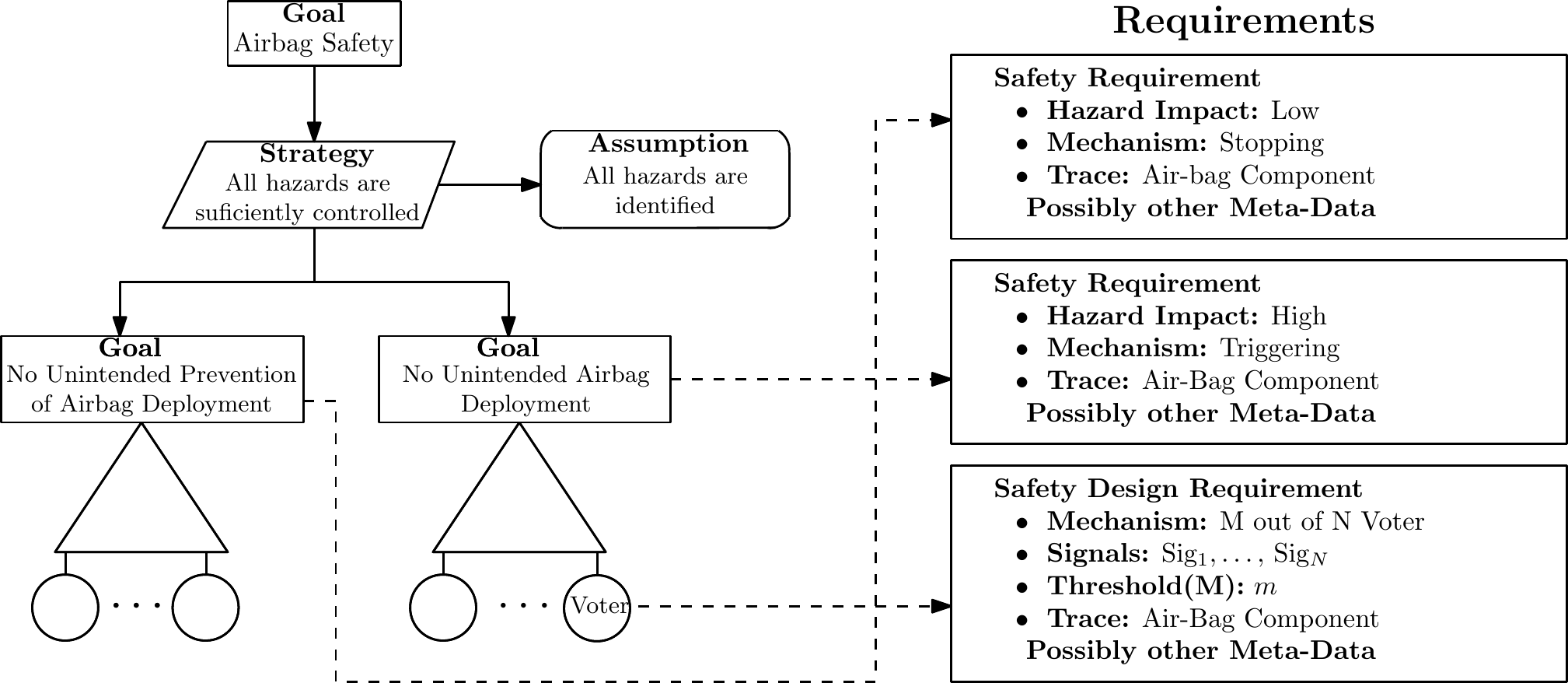}
\end{center}
  \caption{GSN-Model: Airbag Deployment and example of attaching semantics to GSN-Models using domain specific requirements.}
  \label{fig:gns-airbag}
\end{figure*}

We overcome this obstacle by assigning semantics to GSN nodes, inspired by the work on Guide Words (Section~\ref{subsec:safety-security}).\footnote{One could attempt to provide a more general semantics to GSN-Models instead of only its nodes. However, it is not clear yet how this can be done and left to future work. We focus here, instead, on adding enough/minimal meta-data in order to provide useful safety and security integration.} We illustrate this with an example. Consider the GSN-Model depicted to the left in Figure~\ref{fig:gns-airbag} derived from the D\"urrwang \etal's recent work on airbag security~\cite{durrwang18sae}. There are two main safety hazards to be considered for airbag safety: 
\begin{itemize}
  \item \emph{Unintended Prevention of Airbag Deployment}, that is, during a safety critical event, \eg, an accident, the airbag is not deployed. The failure to deploy the airbag reduces the passenger safety during accidents. Notice, however, that other safety mechanisms, \eg, safety belt, may still be enough to ensure passenger safety;

  \item \emph{Unintended Airbag Deployment}, that is, the airbag is deployed in a situation not critical. A passenger, \eg, a kid,  sitting while the car is parked may be hurt if the airbag is deployed. Different from the previous hazard, safety mechanisms, \eg, safety belt, do not ensure the passenger safety. Additional safety mechanisms shall be implemented, such as Voters, as depicted in the GSN-Model.
\end{itemize}

All this information is just described textually in the GSN-Model. However, they shall be reflected in safety and safety design requirements as depicted in Figure~\ref{fig:gns-airbag} by the dashed lines. We propose adding additional meta-data to these requirements, called \emph{domain specific requirements}. The exact meta-data may vary depending on the domain. For embedded systems, safety requirements shall contain data such as:
\begin{itemize}
  \item \emph{Hazard Impact}, which specifies how serious the corresponding hazard is to the item safety. From the reasoning above, the hazard \emph{Unintended Prevention of Airbag Deployment} has a low impact, while 
  \emph{Unintended Airbag Deployment} has a high impact;
  \item \emph{Mechanism} which may be one of the Guide Words detailed in Section~\ref{subsec:safety-security}. For example, the hazard \emph{Unintended Airbag Deployment} is caused by triggering of the air-bag component;
  \item \emph{Trace} from requirement to component is already part of the MBE development. It relates a requirement to a component in the architecture. In this case, the GSN nodes refer to the airbag component.
\end{itemize}

Similarly, solutions, such as voters, are mapped to safety design requirements, for which, we also attach some meta-data. Different types of solutions (FTA, FMEA, Safety Mechanisms) would involve different meta-data. In the case of voters, one specifies the signals used by the voters (Sig$_1, \ldots, $ Sig$_N$), the threshold, $M$, used for deciding when the voter is activated. In our Airbag example, its voter uses signals from the Gyroscope and the Crash detector. Only if all of them indicate a crash situation, then the airbag is deployed. 

Notice that the meta-data attached to domain specific requirements basically reflect the content in the GSN node, but in a uniform format which can be machine-readable. \textbf{This meta-data provides semantic information to GSN nodes.} For example, the meta-data in the requirement associated with the node \emph{No Unintended Airbag Deployment} specifies that the node represents a hazard of high impact which can be the result of triggering the airbag component.\footnote{We are taking extra care to develop domain specific requirements to contain simple, but useful meta-data. While one could be more formal and express requirements in formal languages, such as Linear Temporal Logic~\cite{pnueli77focs}, our experience shows that they are not effective in practice as few engineers and even specialists can write such formulas.}

\begin{figure}
    \begin{center}
  \includegraphics[width=0.95\columnwidth]{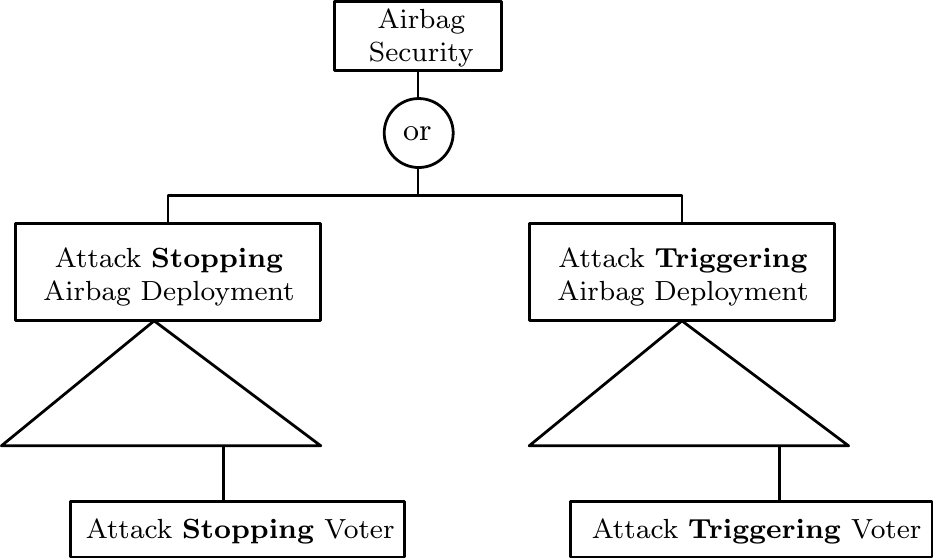}
\end{center}
  \caption{Attack Tree for the Airbag Item.}
  \label{fig:attack-tree-airbag}
\end{figure}

The information associated to GSN-Model is enough to extract useful information for security engineers, allowing to construct (automatically) an attack tree on the security of an item from its corresponding GSN-Model. For example, the attack tree depicted in Figure~\ref{fig:attack-tree-airbag} can be extracted from the airbag GSN-Model depicted in Figure~\ref{fig:gns-airbag}. From the attack tree, security engineers can identify two different types of attacks, stopping airbags or triggering airbag deployment. Notice how the guide words \textbf{stopping} and \textbf{triggering} are used in the construction of the trees. Moreover, from the impact information, security engineers can understand the impact of these attacks, namely, triggering is more harmful than stopping airbag deployment, thus helping prioritize resources, \eg, penetration testing.

Notice that while the voter only appears in one branch of the airbag GSN-Model, attacks appear in both branches of the airbag attack tree. This is because an attack stopping the voter stops airbag deployment. This can be automatically inferred by the meta-data of the voter, first, specifying that it is a M out of N voter and that it is traced to the airbag component.

Solutions, such as, voters, FTA, FMEA, can also be translated to attack sub-trees. 

\begin{itemize}
  \item \textbf{Safety Mechanisms:} 
A safety mechanisms can normally be subject to a large number of attacks as enumerated by Preschern \etal~\cite{preschern13plop}. We can use Guide Words to reduce this list to those attacks that are relevant. For example, an \emph{attack triggering a voter} M out of N can be achieved by \emph{spoofing} M signals or by tampering the voter. It is not necessary to consider denial of service attacks. On the other hand, \emph{stopping the voter} may be achieved by carrying out a denial of service attack on the voter;

  \item \textbf{FTA:} The minimum cut-sets (see Section~\ref{subsec:safety}) resulting from the FTA can be used to construct attack trees. For example, if $\{ev_1, \ldots, ev_n\}$ is a minimum cut-set, then an attack would consist of carrying out attacks to trigger all events $ev_1, \ldots, ev_n$, by, for example, spoofing them; 

  \item \textbf{FMEA:} The table of failures composing an FMEA (see Section~\ref{subsec:safety}) can also be  used to construct an attack tree. In particular, the field failure mode specifies the type of attack on the corresponding function. For example, a \emph{loss of function} entry can be achieved by denying service or tampering the function. Similarly, the severity field indicates how serious the failure mode is and the detection field indicates how disguised the attack can be. It seems possible to transform this information into quantitative information attached to attack trees~\cite{bagnato12ijsse,bistarelli06ares}. This is left for future work.
\end{itemize}

Finally, notice that the attack tree constructed from a GSN-Model provides a preliminary attack tree on the item in question. This tree may be extended considering other possible attacks and attaching counter-measures.

\section{Security to Safety}
\label{sec:sec2saf}
As described in Section~\ref{subsec:safety}, it is possible to attach quantitative evaluation to GSN based on the number of \emph{defeaters} identified and overruled. The result of the quantitative evaluation are three non-negative real values, $B, D, U$, in $[0,1]$ such that $B + D + U = 1$: $B$ is the belief on the GSN-Model, $D$ the disbelief and $U$ the uncertainty. (See the work of Duan \etal~\cite{duan16hases} for more details.) A GSN-Model shall only be acceptable if it has a high enough level of belief and low enough levels of disbelief and uncertainty. The exact degree of belief may depend on different factors, such as, how critical the item. 

Security threats are possible defeaters for GSN-Models as they may invalidate the safety argument. There are the following possibilities according to the security assessment carried out:

\begin{itemize}
  \item \textbf{No Security Assessment for the Item:} If no security assessment has been performed, then it is a defeater that has not yet been outruled and therefore, the uncertainty of the GSN-Model shall be increased. 

  \item \textbf{Existing Security Assessment for the item:} There are two possibilities\footnote{There are many ways to quantify an attack defense tree, \eg, the effort, time, cost required by the attacker to attack an item. Based on these values together with other parameters, \eg, the value of the item, security engineers can evaluate whether the risk is acceptable or not. For example, if all identified attacks to an item take too long to take place, then the risk of such attacks is acceptable.}:

  \begin{itemize}
    \item \textbf{Acceptable Security Risk:} If the security assessment concludes that there is acceptable security risk, that is, identified threats are sufficiently mitigated, then this shall have a positive effect on the belief of the corresponding GSN-Model;

    \item \textbf{Unacceptable Security Risk:} On the other hand, if the identified threats are not sufficiently mitigated, leading to an unacceptable risk, the disbelief of the safety case shall be increased.
  \end{itemize}
\end{itemize}

\begin{figure}
  \begin{center}
    \includegraphics[width=0.7\columnwidth]{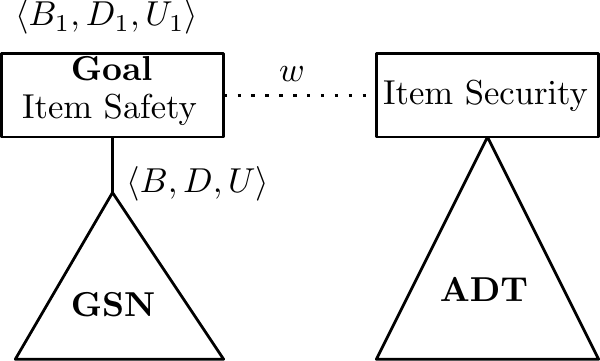}
  \end{center}
  \caption{Illustration of GSN-Model and ADT integration. Here, the values $B,U,D$ are, respectively the levels of belief, disbelief and uncertainty of the safety assessment expressed in the GSN-Model. The new levels of belief, disbelief, and uncertainty, $B_1,U_1,D_1$, are obtained after integrating the security assessments (if any) taking into account the weight $w$, a non-negative number.}
  \label{fig:adt-gsn}
\end{figure}

Figure~\ref{fig:adt-gsn} illustrates how one can integrate GSN-Models and ADTs. The value $w$ is a non-negative value specifying the importance of the security assessment for the item safety. The greater the value of $w$, the greater is the impact of the security assessment. For instance, if $w$ is zero, then the impact of the security assessment on the safety assessment is negligible. Depending on the evaluation of the item security as described above, the levels of confidence of the GSN-Model are updated to $B_1, U_1, D_1$.

We illustrate this with an example implementing a simple update function. Notice, however, that other functions can be used (and subject to future work).
Consider that $B = 0.70, D = 0.20, U = 0.10$ and $w = 2$. The values for belief, disbelief and uncertainty are updated taking into account the security assessment for the item in question if there is any as detailed below and the weight $w$:
  \begin{itemize}
    \item \textbf{No Security Assessment:} In this case, the uncertainty shall increase. We do so by first updating the values for the belief and disbelief, reducing their values according to $w$ as follows:
\[
\begin{array}{l}
    B_1 = B / (1 + w) = 0.7 / (1 + 2) = 0.23 \\
    D_1 = D / (1 + w) = 0.2 / (1  + 2) =   0.07\\
  \end{array}  
\]
Then we compute the uncertainty as follows:
\[
\begin{array}{l}
  U_1 = U + (B - B_1) + (D - D_1) \\ \quad = 0.1 + (0.7 - 0.23) + (0.2 - 0.07) = 0.7.
\end{array}
\]
where the uncertainty increases.

\item \textbf{Acceptable Security Risk:} In this case, the belief shall increase. We do so by carrying out the following computations similar to above, where uncertainty and disbelief decrease:
\[
\begin{array}{l}
    U_1 = U / (1 + w) = 0.1 / (1 + 2) = 0.03 \\
    D_1 = D / (1 + w) = 0.2 / (1  + 2) =   0.07\\
  \end{array}  
\]
Then, we compute the new belief as follows:
\[
\begin{array}{l}
  B_1 = B + (D - D_1) + (U - U_1) \\ \quad = 0.7  + (0.2 - 0.07) + (0.1 - 0.03) = 0.9.
\end{array}
\]
where the belief increases.

\item \textbf{Unacceptable Security Risk:} In this case, the disbelief shall increase. We do so by carrying out the following computations similar to above, where belief and uncertainty decrease:
\[
\begin{array}{l}
    B_1 = B / (1 + w) = 0.7 / (1 + 2) = 0.23 \\
    U_1 = U / (1 + w) = 0.2 / (1  + 2) =   0.07\\
  \end{array}  
\]
Then, we compute the new disbelief as follows:
\[
\begin{array}{l}
  D_1 = D + (B - B_1) + (U - U_1) \\ \quad = 0.2 + (0.7 - 0.23) + (0.1 - 0.03) = 0.7.
\end{array}
\]
where the disbelief increases.
\end{itemize}
Notice that in all cases the new values, $B_1, D_1, U_1$, remain within the interval [0,1] and $B_1+D_1+U_1 = 1$. Moreover, notice that if $w = 0$, then $B_1 = B, D_1 = D, U_1 = U$, that is, the security assessment does not affect the safety assessment.

The use of quantitative evaluations for GSN-Models and ADTs is a way to tackle Challenge 3 (incorporation of relevant security findings into safety assessments) and Challenge 5 (quantitative evaluation for safety including security assessments), as we are able to incorporate the conclusion of security asssessments into the quantitative evaluation of safety assessments and at the same time provide a quantification on the credibility of the safety case in terms of belief, disbelief and uncertainty.

\section{Collaborative Process for Safety and Security}
\label{sec:collaboration}
While this paper's focus is on techniques to integrate security and safety, in this Section, we describe how the techniques described above can be put together as a collaborative process for building an integrated safety and security assessments. We also describe in Subsection~\ref{subsec:saf-sec-processes-challenges} some further challenges that would need to be investigated and sketch ideas on how to tackle these challenges.

In particular, in Sections~\ref{sec:saf2sec} and \ref{sec:sec2saf}, we described how safety assessments in the form of GSN-Models can be used for constructing security assessments in the form of ADT, and moreover, how security assessment results can be integrated into safety assessment by modifying its levels of belief, disbelief and uncertainty. In this section, we describe how these techniques can be put together as a collaborative process for building an integrated safety and security assessments.

\begin{figure}
\begin{center}
  \includegraphics[width=0.99\columnwidth]{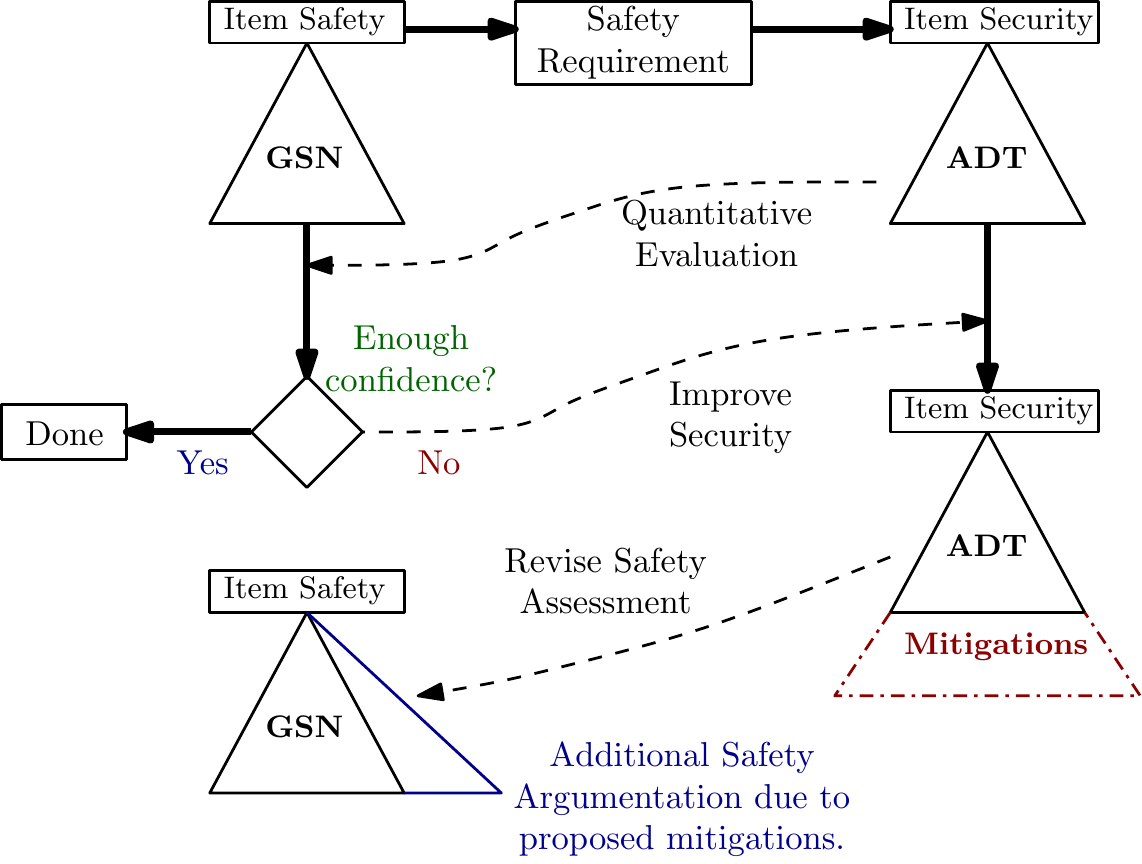}  
\end{center}
\caption{Collaborative Safety and Security Process Cycle.}
\label{fig:collaborative}
\end{figure}

Consider the process cycle illustrated by Figure~\ref{fig:collaborative}. 
\begin{itemize}
  \item \textbf{Initial Safety Assessment:} Assume given an initial safety assesssment for an item represented as a GSN-Model. This 
  starts the process by issuing safety requirements (with meta-data as described in Section~\ref{sec:saf2sec});
  \item \textbf{Security Assessment:} Using the machinery 
described in Section~\ref{sec:saf2sec}, we can build an ADT for the item from the GSN-Model. This ADT may be extended with new threats as well as with mitigation mechanisms;
  \item \textbf{Security Feedback to Safety:} Using the machinery described in Section~\ref{sec:sec2saf}, the evaluation of the security assessment is integrated into the GSN-Model yielding values for belief, disbelief and uncertainty. One finishes the safety and security collaboration if these values are acceptable. Otherwise, there are two possibilities: Either refine the safety assessment, \eg, outrruling more defeaters, or as depicted in Figure~\ref{fig:collaborative}, request for a better security;
  \item \textbf{Additional Mitigations:} Once the request of improving security is received, security engineers can add further mitigation mechanisms in order to improve its security;
  \item \textbf{Safety Revision:} The mitigation mechanisms may impact the safety of the system, \eg, add additional delays or add new single points of failure, etc. This may yield additional safety argumentation. 
\end{itemize}

This collaborative development cycle repeats, possibly adding new safety and security mechanisms, until an acceptable security risk is reached.

\paragraph{Airbag Example:} To illustrate this cycle, let us return to the Airbag safety assessment expressesd by the GSN-Model depicted in Figure~\ref{fig:gns-airbag}. From this GSN-Model, we can construct corresponding attack tree in Figure~\ref{fig:attack-tree-airbag}. This ADT shall yield an unacceptable risk as the threats of stopping the voter and triggering the voter have not been further investigated. This impacts the safety assessment by reducing its belief, disbelief and uncertainty (as described in Section~\ref{sec:sec2saf}). Assume that these values are not acceptable. Thus, the security engineer is requested to improve the ADT. 

In order to improve the ADT, the security engineer may evaluate the risk of, for example, stopping the voter and triggering the voter. However, from the information contained in the hazard meta-data (Figure~\ref{fig:gns-airbag}), the impact of stopping the voter is lower than the impact of triggering the voter. Therefore, the security engineers may decide to further investigate the attack triggering the voter. They may discover, for example, the attack found by Dürrwant \etal~\cite{durrwang18sae} on the security access mechanism which poses a serious threat. In order to mitigate this threat, they can add as counter measure to perform plausibility checks as suggested by Dürrwant \etal~\cite{durrwang18sae}, which would reduce the security risk. 

As new counter-measures have been added, a request to revise safety assessments is issued. Safety engineers have to then argue that the plausibility checks are safe, that is, may not affect the airbag safety, by, for example, preventing airbag deployment. New safety mechanisms may be placed if necessary which may lead to new threats to be analyzed by security engineers. This process ends when the levels of belief, disbelief and uncertainty are acceptable. 

\subsection{Safety and Security Process Challenges}
\label{subsec:saf-sec-processes-challenges}

We identify the following general challenges for any safety and security processes, namely, Incremental Assessment Modifications, Assessment Completeness and Assessment Verification and Validation:

   \paragraph{Incremental Assessment Modifications:} In order to be practical, \emph{incremental changes} to the system shall require only \emph{incremental changes} to the existing assessments. That is, whenever there is a small modification to the system, assessments do not need to be re-done from scratch, but only small focused modifications are needed that are relevant to the changes made. Of course, the definition of \emph{incremental} depends on the system in question. It may be, for example, the inclusion of a localized counter-measure or safety solution in a system. 

   Indeed, as described above, the inclusions of plausibility checks as mitigation mechanisms for attacks on the Airbag system impacts the safety assessment, but only in an incremental fashion: The safety engineer shall evaluate whether these plausibility checks can increase the chance of the hazard of stopping airbag deployment when it shall be deployed. The other concerns in the assessment do not need re-assessment.

   This example also illustrates how Guide Words and models can be used for identifying which parts of assessments are impacted by incremental changes to the system. Mitigations added for mitigating attacks related to component hazards associated with the Guide Word ``trigger'' shall require further safety assessments on component hazards associated to the Guide Word ``stop''.  We believe that further generalizations are possible with the other Guide Words.

\paragraph{Assessment Completeness:} A safety and security process shall produce assessments that are complete, that is, covers all hazards and possible threats. For safety there are precise techniques, such as FTA, FMEAs, for assessing the completeness of safety assessments with respect to safety hazards. These are, in practice, enforced by certification agencies using, for example, ASIL or DAL levels, requiring \emph{precise probability of failure requirements}. This is less so for security assessments, as there is no precise definition of when the risk is acceptable. 

   Since when integrating safety and security, the confidence on safety assessment depends on evaluation of the security assessment, the confidence of safety assessments are no longer precise by just using ASIL levels. One needs, therefore, definitions of completeness of security assessments. 

   We elaborate on some possible definitions of security assessment completeness. We also envision definitions that combine these notions of completeness depending on the type of system.

\bigskip

   \begin{itemize}
     \item \textbf{Complete with respect to some formal property:} Given a formal property, such as a set of behaviors\footnote{Or even sets of sets of behaviors, as in hyper-properties~\cite{clarkson10jcs}.}, the assessment is complete when it follows that the system in question satisfies the formal property. The main advantage of this definition of completeness is its formal nature. However, this is also its main disadvantage as few engineers are able to write such properties and moreover, they are many times hard to validate. This is either because existing automated methods, such as, Model-Checkers, do not scale, or semi-automated methods require a great deal of expertise;

     \item \textbf{Complete with respect to Intruder Capabilities:} Completeness can depend on the assumptions on the intruder capabilities, such as, whether he can inject messages, replay messages, block messages. That is, the security assessment is said to be complete if the system in question is secure with respect to the given assumptions of the intruder capabilities.
     Such intruder models have been successfully used, for example, in protocol security verification. One advantage is that it can also be made more precise using formal definitions. One disadvantage is that it has been shown, until now, to work only for logical attacks. There are less success stories for other types of attacks, such as attacks on cyber-physical domains, where there intruders may use, for example, side-channels;

     \item \textbf{Complete with respect to Known Attacks:} Given a set of known attacks, the security assessment is complete when it follows that the system in question has acceptable risk with respect to each known attack. An advantage of this approach is that it is attack-centric, focusing on concrete attacks. However, a disadvantage is that the size of the set of known attacks can increase rapidly. Moreover, in order to validate the assessment, security engineers would need to reproduce each attacks and if it is discovered that the system is vulnerable to an attack, possible causes shall be identified, which may take much effort being even impractical;

     \item \textbf{Complete with respect to Known Defects:} Given a set of known defects, such as vulnerabilities, the assessment is complete if it demonstrates that the system in question does not have anyone of the given defects. A main advantage with respect to, for example, the definition of completeness with respect to known attacks, is that, in principle, the number of defects is less than then the number of attacks, as different attacks may exploit the same defect. For example, a number of attacks (Buffer-Overflows, SQL injections) exploit the lack of string input checks.  Moreover, there are methodologies, such as, static analysis, defect-based testing, that can be used to identify and mitigate many defects.
   \end{itemize}

\paragraph{Assessment Verification and Validation (V\&V):} The assessment shall also contain enough information for engineers to carry out verification and validation plans. At the end, there shall be guarantees that the system in question satisfies the properties the assessment claims to establish, namely, that the system in question has acceptable safety and security risks.

In particular, the assessment shall contain information on what types of validation and verification plans shall be deployed. This is closely related to the notion of completeness of the assessment. For example, assessments based on defects can be verified and validated by techniques, such as, defect-based testing. On the other hand, assessments based on particular attacks shall describe how to carry out these attacks.

\section{Trade-off Analysis}
\label{sec:trade-off}
In this section, we describe methods towards carrying out trade-off analysis between safety and security mechanisms. Such analysis may help decide which mechanisms to be implemented. It may   be that there are synergies between safety and security mechanisms which would make them redundant. For example~\cite{glas14safesec,novak07etfa}, CRC checks used for checking the integrity of messages and MAC used to ensure that no message is corrupted. Therefore, MAC could replace CRC, rendering CRC not needed. On the the other hand, safety and security mechanisms may conflict, that is, interfere with their purposes. In such cases, one may have to decide on alternative mechanisms or ultimately choose either safety or security.

\begin{figure}
\begin{center}
  \includegraphics[width=1\columnwidth]{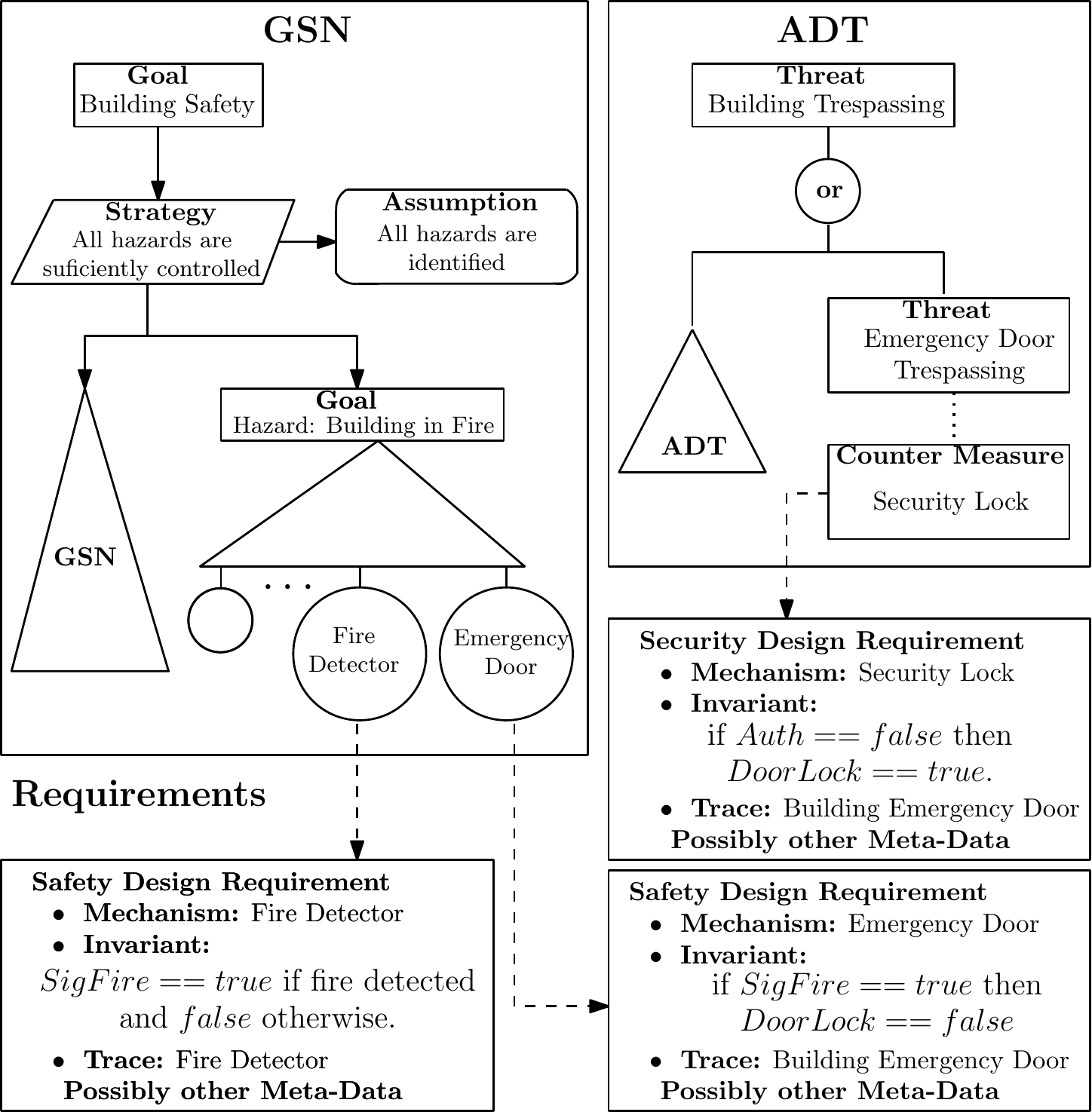}  
\end{center}
\caption{Illustration of GSN-Model and ADT for detecting conflicts on proposed safety and security mechanisms.}
\label{fig:gsn-adt-building}
\end{figure}

We illustrate how MBE can be used to carry out these trade-offs with an example. 
Consider the safety and security arguments expressed by a GSN-Model and an ADT and depicted in Figure~\ref{fig:gsn-adt-building} of a building. The arguments express the following concerns: 

\begin{itemize}
  \item \textbf{Building Safety:} The GSN-Model specifies, among other possible hazards, controlling the hazard of people getting hurt when the building is in fire. It proposes as solutions installing a \emph{fire detector} and an \emph{emergency door}. These solutions are associated with domain specific safety requirements (pointed by the corresponding dashed lines). These requirements are functional requirements specifying that the \emph{boolean signal} $SigFire$ is true when a fire is detected and false otherwise and that when $SigFire$ is true then the emergency door shall be unlocked, that is, the signal $DoorLock$ is false. 

\item \textbf{Building Trespassing:} The ADT breaks down the threat of a malicious intruder trespassing in the building. A possibility is by entering the building using the emergency door. This threat is mitigated by installing a security lock in the emergency door with an \emph{authentication mechanism} (\eg, biometric, code, card). This mitigation is associated to a domain specific security requirement, specifying the function of the security door: If the authentication mechanism signal $Auth$ is false, then the emergency door shall be locked.
\end{itemize}

Given these arguments (GSN-Model and ADT) and its associated domain specific requirements, it is possible to identify potential conflicts: one simply needs to check whether the requirements have intersecting set of signal names. In this example, the $DoorLock$ output signal is mentioned in both the security lock requirement and in the emergency door requirement. A priori, the fact that the same signals are mentioned does not mean that there is a conflict, but only that these are \emph{potential candidates for conflicts}. This is just one possible method for identifying conflict candidates. Other methods may use the trace, the type of requirements, etc. It is important, however, to have simple mechanisms to determine these candidates as in a usual development a large number of requirements are specified.

Once the potential candidates are identified, it remains to check whether they are indeed conflicting. We illustrate how this can be done using off-the-shelf tools. First, we extract the logical clauses in the requirements, where $SigFire$ and $Auth$ are input signals and $DoorLock$ is the output signal:
\[
    \begin{array}{ll}
      SigFire \Rightarrow \neg DoorLock & \textrm{ from emer. door req.}\\
      \neg Auth \Rightarrow DoorLock & \textrm{ from sec. lock req.}
    \end{array}
\]
The question is whether these clauses can entail contradicting facts. This is clearly the case as when $SigFire$ is true implies that $DoorLock$ is false, and when $Auth$ is false implies that $DoorLock$ is true, thus yielding a contradiction. 

For such (more or less) simple specifications, one can detect such 
contradiction manually. However, as specification become more complicated and the number of requirements increase, checking all potential contradictions for actual contradictions becomes impractical. It is much better is to automate this process as we demonstrate with this example.

Before, however, we should point out that traditional classical logic (propositional logic) is not suitable for this problem because of the famous frame problem~\cite{McCHay69}. This is because only propositions that are supported by the extracted logical clauses shall be derivable. One way to solve this problem is to use the \emph{Closed World Assumption}~\cite{lifschitz85ai} from the knowledge representation literature~\cite{baral.book}. We will use here the logic paradigm Answer-Set Programming (ASP)~\cite{gelfond90iclp,baral.book} which allows specifications using the Closed World Assumption and the solver DLV~\cite{leone06tcl}\footnote{\url{http://www.dlvsystem.com/}} which supports ASP. 

We start by adding for each predicate ($SigFire$, $DoorLock$, $Auth$), a \emph{fresh predicate} with a prefix $neg$ corresponding to its classical negation ($negSigFire$, $negDoorLoc$, $negAuth$). Thus it should not be possible that, for instance, $negDoorLock$ and $DoorLock$ are both inferred from the specification at the same time as this would be a contradiction. 
Second, we translate the clauses above into the following ASP program using DLV syntax:\footnote{A logic programming clause of the form \texttt{A :- B1, ..., Bn} shall be interpreted as the clause $B_1, \ldots, B_n \Rightarrow A$.}
\begin{verbatim}
1: DoorLock :- negAuth.
2: negDoorLock :- SigFire.
3: negAuth v Auth.
4: negSigFire v SigFire.
5: contradiction :- DoorLock, negDoorLock.
6: :- not contradiction.
\end{verbatim}
The first two lines are direct translations of the clauses above. The lines 3 and 4 specify, respectively, that either $negAuth$ or $Auth$ is true\footnote{The symbol $\texttt{v}$ should not be interpreted as ``or'', but more close to ``x-or'', though not exactly. More details can be found at~\cite{leone06tcl}.} and either $negSigFire$ or $SigFire$ is true.  Line 5 specifies that there is a $contradiction$ if both $DoorLock$ and $negDoorLock$ can be derived. Finally, line 6 is a constraint specifying that only solutions that contain $contradiction$ shall be considered. This is because for this example we are only interested in finding (logical) contradictions. If there is no such solution, then the theory is always consistent and therefore, the requirements are not contradicting. 

For the program above, however, we obtain a single solution (answer-set) when running this program in DLV:
\begin{verbatim}
   {DoorLock, negAuth, negDoorLock, 
              SigFire, contradiction}
\end{verbatim}  
indicating the existence of a contradiction, namely, the one we expected where $Auth$ is false and $SigFire$ is true.

Once such contradictions are found, safety and security engineers have to modify their requirements. A possible solution is for the security lock to check whether there is a fire or not, that is, having the following invariant:
\begin{center}
  if $Auth == false$ and $SigFire == false$ then $DoorLock == true$.
\end{center}
which resolves the contradiction as can be checked by again using DLV.

Sun \etal~\cite{sun.unpublished} propose determining such conflicts by using the rewriting tool Maude~\cite{clavel-etal-07maudebook}. While their encoding is much more involved than ours, the use of Maude has the potential of finding different types of conflicts, such as involving delays. This is because the encoding in Maude specifies part of the operational semantics of the system, while our encoding only takes into account the logical entailment.

Finally, this proof-of-concept example illustrates how conflicts are detected. It seems possible to also determine when requirements support each other, by adding suitable meta-data in domain specific requirements. For example, CRC and MAC solutions for the same communication channels. 
Further investigation is left for future work.



\section{Conclusions}
\label{sec:steps}

The main goal of this white paper is to set the stage for Safety and Security Engineering. We identified some key technical challenges  in Section~\ref{sec:challenges}. We then illustrated with examples techniques that can help address some of these challenges. For example, we showed how to extract security relevant information of safety assessments by translating GSN-Models into ADTs. For this, we provided semantics to GSN nodes by using \emph{domain-specific requirements}. We also showed how to use existing machinery on quantitative evaluation of GSN-Model and ADTs to incorporate the findings of security assessments into safety assessments. We then proposed a collaborative development process where safety and security engineers incrementally build arguments using their own methods. Finally, we demonstrated how paradigms, such as Answer-Set Programming, can be used to identify when safety and security assessments are conflicting.

This is clearly the start of a long and interesting journey towards Safety and Security Engineering. Besides further developing the techniques illustrated in this white paper, we identify the following future work categorized into Techniques, Processes and Certification:

\begin{itemize}

  \item \textbf{Techniques:} As pointed out throughout the white paper, the techniques we illustrate are going to be subject of intensive future work. We would like to answer questions such as: which meta-data should be added to domain-specific requirements or to models in order to enable further automated model translation? How can different security domains impact safety cases? How can we automatically detect other types of contradictios, such as timing contradictions? Finally, how can trade-off analysis be compiled so to facilitate conflict solving?

  \item \textbf{Collaborative Processes:} While here we illustrate a collaborative process involving safety and security concerns only, we are investigating how to extend this collaboration with other aspects, such as performance and quality. We are also investigating how better tooling can make the collaborative process go smoothly, \eg, automated notifications. Also we are investigating techniques for addressing the challenges described in Section~\ref{subsec:saf-sec-processes-challenges}.

  \item \textbf{Certification:} We are currently investigating how the techniques and the collaborative process cycle relate with certifications, such as the ISO 26262~\cite{iso26262}. A particular goal for future work is to build automated techniques that can be used to support the building of convincing safety and security assessments, complementing recent work~\cite{sabaliauskaite18ijas} on the topic.

\end{itemize}

Finally, we plan to apply the techniques in extended use-cases from different domains. We will also report these results as scientific papers and technical reports to industry.



\phantomsection
\section*{Acknowledgments} 
We thank our industry partners, in particular, Airbus Defense and BMW Research, for valuable discussions. We also thank the AF3 team for helping us with the implementation of features in AF3. Finally, we also thank the fortiss Safety and Security Reading group.


\phantomsection
\bibliographystyle{plain}
\balance
\bibliography{bib}


\end{document}